\documentclass{article}
\usepackage{makeidx}
\usepackage{amssymb}
\usepackage{amsfonts}
\usepackage{amsthm}
\usepackage{amsmath}
\usepackage{stmaryrd}
\usepackage{graphicx}
\usepackage[all]{xy}
\newcommand{\machine}{\mathit{H}}
\newcommand{\memory}{\mathit{M}}
\newcommand{\baseset}{\mathit{B}}
\newcommand{\sts}{\mathcal{S}}
\newcommand{\state}{\mathit{S}}
\newcommand{\opns}{\mathcal{O}}
\newcommand{\operation}{\mathit{O}}
\newcommand{\bactions}{\mathit{A}}
\newcommand{\bas}{\mathcal{A}}
\newcommand{\baif}[1]{\llbracket #1 \rrbracket}
\newcommand{\tr}{\mathsf{T}}
\newcommand{\fa}{\mathsf{F}}
\newcommand{\aply}{\bullet_\machine}
\newcommand{\fine}{\mathcal{E}_\mathbf{fin}}
\newcommand{\frterms}{\mathcal{T}_\mathbf{finrec}}
\newcommand{\bidm}{\bactions'}
\newcommand{\rr}{\mathsf{rr}}
\newcommand{\loa}{\mathsf{load}}
\newcommand{\sto}{\mathsf{store}}
\newcommand{\blset}{\{\tr,\fa\}}
\newcommand{\misa}{\mathcal{MISA}_\mathrm{sls}}
\newcommand{\mdata}{\mathsf{M}_\mathsf{data}}
\newcommand{\baddr}{\mathsf{B}_\mathsf{addr}}
\newcommand{\bdata}{\mathsf{B}_\mathsf{data}}
\newcommand{\fmdata}{\mathsf{m}_\mathsf{data}}
\newcommand{\mou}{\mathsf{M}_\mathsf{ou}}
\newcommand{\bit}{\mathsf{Bit}}
\newcommand{\mld}{\mathsf{M}_\mathsf{ld}}
\newcommand{\msd}{\mathsf{M}_\mathsf{sd}}
\newcommand{\mla}{\mathsf{M}_\mathsf{la}}
\newcommand{\msa}{\mathsf{M}_\mathsf{sa}}
\newcommand{\fmld}{\mathsf{m}_\mathsf{ld}}
\newcommand{\fmsd}{\mathsf{m}_\mathsf{sd}}
\newcommand{\fmla}{\mathsf{m}_\mathsf{la}}
\newcommand{\fmsa}{\mathsf{m}_\mathsf{sa}}
\newcommand{\setu}{\mathsf{B}_\mathsf{load}}
\newcommand{\setv}{\mathsf{B}_\mathsf{store}}
\newcommand{\tpfc}{\mathcal{TPFC}}
\newcommand{\sdata}{\mathsf{S}_\mathsf{data}}
\newcommand{\tdata}{\mathsf{T}_\mathsf{data}}
\newcommand{\tp}{\mathit{T}}
\newcommand{\bta}{\ensuremath{\mbox{BTA}}}
\newcommand{\bppa}{\ensuremath{\mbox{BPPA}}}
\newcommand{\di}{\mathsf{D}}
\newcommand{\st}{\mathsf{S}}

\newcommand{\sltstp}{\mathsf{tau}}
\newcommand{\pcc}[1]{\unlhd #1 \unrhd}
\newcommand{\aslp}{\bas_\sltstp}
\newcommand{\sol}[1]{\langle{#1}{\rangle}}
\newcommand{\tone}{\mathrm{T1}}
\newcommand{\rdp}{\mathrm{RDP}}
\newcommand{\rsp}{\mathrm{RSP}}
\newcommand{\btarec}{\bta + \ensuremath{\mbox{REC}}}

\newtheorem{lem}{Lemma}
\newtheorem{defn}{Definition}
\newtheorem{thm}{Theorem}
\newtheorem{cor}{Corollary}
\begin{document}
\title{On Transformations of Load-Store Maurer Instruction Set Architectures}
\author{Tie Hou\\
\,Informatics Institute, University of Amsterdam,\\ Kruislaan 403, 1098 SJ Amsterdam, The Netherlands\\
%\texttt{hou@science.uva.nl}
}
\maketitle
%
%%%%%%%%%%%%%%%%%%%%%%%%%%%%%%%%%%%
\section{Introduction}
Maurer proposes a model for computers from the viewpoint of general
function and set theory in~\cite{Mau:tci:1966, Mau:tci:2006}.
Mathematical machines (Turing machines, push-down automata,
etc.) are widely known for their inadequate representation of modern
computers, but Maurer's model gives a leading solution. \emph{Maurer
machines}~\cite{BM:mcsc:2007}, introduced by Bergstra and
Middelburg, are based on this model and basic thread algebra with
the operator for applying threads to Maurer machines. \emph{Basic
thread algebra} ($\bta$), which was introduced as \emph{Basic
Polarized Process Algebra ($\bppa$)} in~\cite{BL:pasc:2002}, is a
theory that describes the behaviour of deterministic sequential
programs under execution. The behaviours concerned are supposed to
be \emph{threads} in $\bta$ (see more in~\cite{BM:ta4si:2007}).

In \emph{load-store} (or \emph{register-register}) architectures
(see, e.g.,~\cite{HP:ca:2003}), we have explicit instructions that
access memory only. Load instructions read data from the memory and
copy them to registers. Store instructions write data from registers
to the memory. Computers of today use load-store architectures,
because (1) register access is faster than memory access; (2)
registers allow for compiler optimisations, e.g., an expression may
be evaluated in any order of execution; (3) registers can be used to
hold all the variables relevant for a specific code segment, so the
operations are faster.

In~\cite{BM:ousl:2007}, Bergstra and Middelburg introduced the
concept of a \emph{strict load-store Maurer instruction set
architecture} (strict load-store Maurer ISA, for short) and studied
under what conditions and how these conditions can affect the
transformations on the states of the memory of a strict load-store
Maurer ISA to be achieved.

There are mainly three parts in a load-store instruction set
architecture: a memory that contains data, registers, and an
operating unit that processes data. In this paper, we study how
certain conditions can affect the transformations, when half of the
data memory serves as the part of the operating unit.

The rest of the paper is organised as follows. First of all, we
review basic thread algebra and Maurer machines in
Section~\ref{sectbta} and Section~\ref{sectmm}, respectively. Next,
in Section~\ref{sectapp}, we describe the notion of the apply
operator. Following this, we explain the strict load-store
instruction set architectures in Section~\ref{secisa}. After that,
in Section~\ref{secttpf}, we review the concept of thread powered
function classes and show two results of the completeness. Then we recall
an incompleteness in Section~\ref{sectinc}. Finally, we give some
concluding remarks in Section~\ref{sectcon}.
%
%%%%%%%%%%%%%%%%%%%%%%%%%%%%%%%%%%%
\section{Basic Thread Algebra}\label{sectbta}
Consider a fixed but arbitrary finite set $\bas$ of \emph{basic
actions} with $\sltstp\notin\bas$. We denote $\bas\cup\{\sltstp\}$
by $\aslp$. The signature of $\bta$ consists of the following
constants and operators:
\begin{enumerate}
\item the \emph{deadlock} constant $\di$; \item the
\emph{termination} constant $\st$; \item for each $a\in \aslp$, a
binary \emph{postconditional composition} operator $\_\pcc{a}\_$.
\end{enumerate}
With $\di$ an inactive behavior is indicated and with $\st$ a
successful terminating behavior is denoted. A single action is not
a thread, and finite threads always end in $\st$ or $\di$. The thread
$x\pcc{a} y$ will first perform $a$ and then proceed as $x$ if the
processing of $a$ produces the \emph{positive} reply $\tr$, and it
will proceed as $y$ if the processing of $a$ produces the
\emph{negative} reply $\fa$. We abbreviate $P\pcc{a} P$ using the
\emph{action prefixing} operator: $a\circ P$ and take $\circ$ to
bind strongest. The action $\sltstp$ will always produce a
positive reply. The axiom for this action is given in
Table~\ref{axm4bta}. Using the action prefixing operator, axiom
$T1$ can be also written for short as: $x\pcc{\sltstp}
y=\sltstp\circ x$.
\begin{table}[!ht]\centering
\caption{Axioms for $\bta$}\label{axm4bta}
\begin{tabular}{@{} lc @{}}
\hline $x\pcc{\sltstp} y=x\pcc{\sltstp}x$ & $\tone$\\
 \hline
\end{tabular}
\end{table}

Every thread in $\bta$ is finite in the sense that the number of
consecutive actions it can perform is bounded. Infinite threads
can be defined using guarded recursive specifications.

A \emph{guarded recursive specification} over $BTA$ is a set of
recursion equations $\{X_i=t_i(X)|X_i\in V_E\}$, where
$V_E=\{X_1,X_2,\ldots,X_n\}$ is a set of all variables that occur
on the left-hand side of an equation in $E$, $X$ is a vector
containing all variables in $V_E$, i.e. $X=X_1,\ldots,X_n$, and
$t_i$ is a term of the form $\di$,$\st$ or $t\pcc{a} t'$ ($t$ and
$t'$ are terms of $\bta$ that contain only variables from $X$).

A \emph{solution} for a recursive equation is a thread that solves
the equation. We use the constant $\sol{X_i|E}$ to denote the
solution for the recursive equation $(X_i=t_i(X))\in E$. A
solution for a guarded recursive specification $E$, with
$V_E=\{X_1,\ldots,X_n\}$, is a vector
$\sol{X_1|E},\ldots,\sol{X_n|E}$ such that substituting each
variable in $V_E$ by its respective solution turns all equations
in $E$ into true statements. Once $E$ is declared, $\sol{X_i|E}$
can be abbreviated by $\sol{X_i}$. We give the axioms for guarded
recursion in Table~\ref{axm4rdp}.
\begin{table}[!ht]\centering
\caption{Axioms for guarded recursion}\label{axm4rdp}
\begin{tabular}{@{} lc @{}}
\hline $\sol{X_i|E}=t_i(\sol{X_1|E},\ldots,\sol{X_n|E})$ $(i\in\{1,\ldots,n\})$ & $\rdp$\\
$E\Rightarrow X_i=\sol{X_i|E}$ & $\rsp$\\
 \hline
\end{tabular}
\end{table}
The \emph{recursive definition principle} ($\rdp$) states that
$\sol{X_1|E},\ldots,\sol{X_n|E}$ is a solution for $E$. The
\emph{recursive specification principle} ($\rsp$) states that this
solution is the only one.

We write $\btarec$ for $\bta$ extended with the constants for
solutions of guarded recursive specifications and axioms $\rdp$ and
$\rsp$.

From now on, we write $\fine(\bactions)$, where
$\bactions\subseteq\bas$, for the set of all finite guarded
recursive specifications over $\bta$ that contain only
postconditional operators $\_\pcc{a}\_$ for which $a$ ranges over
$\bactions$, and $\frterms(\bactions)$, where
$\bactions\subseteq\bas$, for the set of all closed terms of
$\btarec$ that contain only postconditional operators $\_\pcc{a}\_$
for which $a$ ranges over $\bactions$ and only constants
$\sol{X_i|E}$ for which $E$ ranges over $\fine(\bactions)$.

We give the following definition of the set of thread states, which
will be used later in Section~\ref{secttpf}.
\begin{defn}
Let $\bas$ be some model of $\btarec$, and let $p$ be an
element from the domain of $\bas$. Then the set of \emph{states} of
$p$, written $Res(p)$, is inductively defined as follows:
\begin{enumerate}
\item $p\in Res(p)$;
\item if $q\pcc{a} r\in Res(p)$, then $q,r\in Res(p)$.
\end{enumerate}
\end{defn}

In subsequent sections, the following threads, which have more than one initial states, are not used.
\begin{figure}[!ht]
\begin{center}
$\begin{array}{c} 
\xymatrix{
a\circ\st \ar[dr]_{a} && b\circ\st \ar[dl]^{b}\\
&\st&
}
\end{array}$
\end{center}
\caption{Connected Thread}
\label{conthd}
\end{figure}
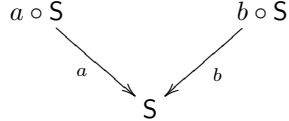
%
%%%%%%%%%%%%%%%%%%%%%%%%%%%%%%%%%%%
\section{Maurer Machines}\label{sectmm}
In this section we review Maurer machines, which were first
introduced in~\cite{BM:mcsc:2007}.

Most modern computers use the binary system, i.e., information is
exchanged and processed internally using 2 as numerical base.
Theoretically we can also use any number as the base, such as 3, 5, 8, etc. Therefore, a
computer can be constructed to the base $n$, which means that
information is virtually operated using only the digits from $0$
through $n-1$. We assume that the base $n$ is constant over the
whole computer.

Every computer has a \emph{memory}. We represent the memory of a
computer as a set $\memory$. Registers are regarded as subsets of
$\memory$. We consider a set $\baseset$ as the \emph{base set},
whose cardinality is the \emph{base} of the computer. If the base
of a computer is $n$, the base set of this computer is the set of
all integers from $0$ to $n-1$. A \emph{state} of the computer is
represented as an arbitrary map from $\memory$ to $\baseset$. We
can change one state to another by performing \emph{operations}.

Maurer machines are based on this simple model of
computers. The memory of a Maurer machine consists of memory
elements. Every memory element contains a value from the base set
of the Maurer machine as a content. The contents of all memory
elements build up a state of the Maurer machine. The Maurer machine
processes a basic action by performing the operation associated
with the basic action. The execution of an operation carries out
the passing from one state to the next. As a result of state
changes, the content of the memory element associated with the
basic action is changed to the reply produced by the Maurer
machine.

Now we give the following definition of a Maurer machine.
\begin{defn}
Let $\memory$ be a non-empty set, let $\baseset$ be a set with
$card(\baseset)\ge 2$ (which means $\baseset$ contains at least two
members $\tr$ and $\fa$), let $\sts$ be a set of functions $\state$:
$\memory\rightarrow \baseset$, let $\opns$ be a set of functions
$\operation$: $\sts\rightarrow \sts$, let $\bactions\subseteq \bas$
be a set, let $\baif{\_}$:
$\bactions\rightarrow(\opns\times\memory)$ be a function, satisfying
the following conditions:
\begin{enumerate}
\item [-] if $S_1,S_2\in\sts$, $M'\subseteq\memory$, and
$S_3$:$\memory\rightarrow \baseset$ is such that $S_3(x)=S_1(x)$
if $x\in M'$ and $S_3(x)=S_2(x)$ if $x\notin M'$, then
$S_3\in\sts$; \item [-] if $S_1,S_2\in\sts$, then the set $\{x\in
M\mid S_1(x)\not = S_2(x)\}$ is finite; \item [-] if $S\in\sts$,
$a\in\bactions$, and $\baif{a}=(\operation,m)$, then $S(m)\in
\{\tr,\fa\}$.
\end{enumerate}
Then the 6-tuple
$\machine=(\memory,\baseset,\sts,\opns,\bactions,\baif{\_})$ is a
\emph{Maurer machine}. The set $\memory$ is the \emph{memory} of
$\machine$; the set $\baseset$ is the \emph{base set} of
$\machine$; the members of $\sts$ are the \emph{states} of
$\machine$; the members of $\opns$ are the \emph{operations} of
$\machine$; the members of $\bactions$ are the \emph{basic
actions} of $\machine$; and the function $\baif{\_}$ is the
\emph{basic action interpretation function} of $\machine$.
\end{defn}

Every operation $\operation:\sts\rightarrow\sts$ is associated
with two subsets of $\memory$. For example, if we want to move the
data in the memory $Y$ to the register $R$, we are implying
$Y$ and $R$ are proper subsets of $\memory$. We give the relation
between $\operation$ and these two subsets by the following
notions of input and output regions of an operation, which will be
used later in Section~\ref{secisa}.
\begin{defn}
Let $\machine=(\memory,\baseset,\sts,\opns,\bactions,\baif{\_})$
be a Maurer machine, and let $\operation:\sts\rightarrow\sts$.
Then we define the \emph{input region} of $\operation$, written
$IR(\operation)$, and the \emph{output region} of $\operation$,
written $OR(\operation)$, which are the subsets of $\memory$, as
follows:
\begin{eqnarray*}
IR(\operation)&=&\{x\in\memory\mid\exists\state_1,\state_2\in\sts.(\forall z\in\memory\backslash\{x\}.\state_1(z)=\state_2(z)\land \\
&&\qquad \quad\exists y\in OR(\operation).\operation(\state_1)(y)\not =\operation(\state_2)(y))\}, \\
OR(\operation)&=&\{x\in\memory\mid\exists\state\in\sts.\state(x)\not =\operation(\state)(x)\}.
\end{eqnarray*}
\end{defn}

According to this definition, in the above example, we call $Y$
the input region and $R$ the output region of $\operation$. Each
operation takes data only from its input region and places data
only in its output region.
%
%%%%%%%%%%%%%%%%%%%%%%%%%%%%%%%%%%%
\section{Application of Threads to Maurer Machines}\label{sectapp}
The binary apply operator $\_\aply\_$ connects a thread and a
state of a Maurer machine, and yields either a state of the Maurer
machine or the \emph{undefined state} $\uparrow$. In other words,
$p\aply \state$ indicates the resulting state after the Maurer
machine
$\machine=(\memory,\baseset,\sts,\opns,\bactions,\baif{\_})$
executes all the basic actions performed by the thread
$p\in\frterms(\bactions)$ from the initial state $\state\in\sts$.
Let $(\operation_a,m_a)=\baif{a}$ for all $a\in\bactions$.
$\machine$ executes a basic action $a$ by performing
$\operation_a$. This leads to a state change. In the
resulting state, the reply produced by $\machine$ is the content
in $m_a$. If $p$ is $\st$, no state changes. If $p$ is $\di$, the
result is $\uparrow$.

Then we give the following defining equations for the apply
operator in Table~\ref{aplydef}, where $a$ ranges over
$\bactions$, and $\state$ ranges over $\sts$.
\begin{table}[!ht]\centering
\caption{Defining equations for apply operator}\label{aplydef}
\begin{tabular}{@{} lr @{}}
\hline $x\aply\uparrow=\uparrow$ &\\
$\st\aply\state=\state$ &\\
$\di\aply\state=\uparrow$ &\\
$(x\pcc{a} y)\aply\state=x\aply\operation_a (\state)$ & if
$\operation_a(\state)(m_a)=\tr$\\
$(x\pcc{a} y)\aply\state=y\aply\operation_a (\state)$ & if
$\operation_a(\state)(m_a)=\fa$\\
 \hline
\end{tabular}
\end{table}
%
%%%%%%%%%%%%%%%%%%%%%%%%%%%%%%%%%%%
\section{Strict Load-Store Maurer ISAs}
\label{secisa}
In this section we review a strict load-store Maurer
ISA~\cite{BM:ousl:2007,BM:mcpi:2008}.

\begin{figure}[!ht]
\centering
\includegraphics{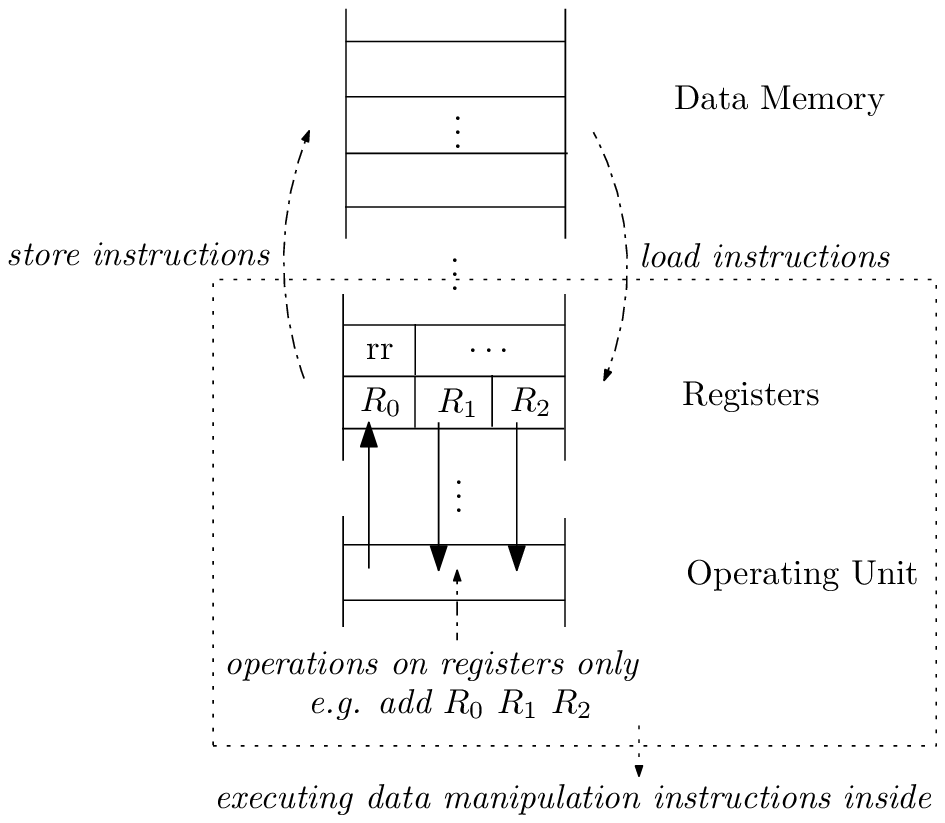}
\caption{Strict Load-Store Maurer ISA} \label{figlsa}
\end{figure}

The basic idea of a strict load-store Maurer ISA is the
following: in the setting of Maurer machines, a segmented memory
is used as a main memory to contain data, and a small segmented
memory is used as an operating unit to process data, as shown in Figure~\ref{figlsa}.
Only load and store instructions can access the data memory, moving
data from the memory to the register, or to the memory from the
register, respectively. All other instructions (e.g., instructions
for data manipulation) can use only register operands. Operations
(such as, calculating a data address, add, subtraction, AND, shifts,
etc.), taking operands from registers, are executed in the operating
unit. The result is stored back to a register. Without loss of
generality, we assume that data is restricted to the natural
numbers.

A strict load-store Maurer ISA has the following parameters:
\begin{enumerate}
\item [-] an address width $k$; \item [-] a word length $l$; \item
[-] a bit size $m$ of the operating unit; \item [-] a number $u$
of pairs of address and data registers for load instructions;
\item [-] a number $v$ of pairs of address and data registers for
store instructions; \item [-] a set $\bidm$ of basic instructions
for data manipulation.
\end{enumerate}
The symbols can be regarded as follows:
\begin{enumerate}
\item [-] $k$: the number of bits used for the binary
representation of addresses of data memory elements; \item [-]
$l$: the number of bits used to represent data in data memory
elements; \item [-] $m$: the number of bits that the internal
memory of the operating unit contains.
\end{enumerate}

\begin{figure}[!ht]
\centering
\includegraphics{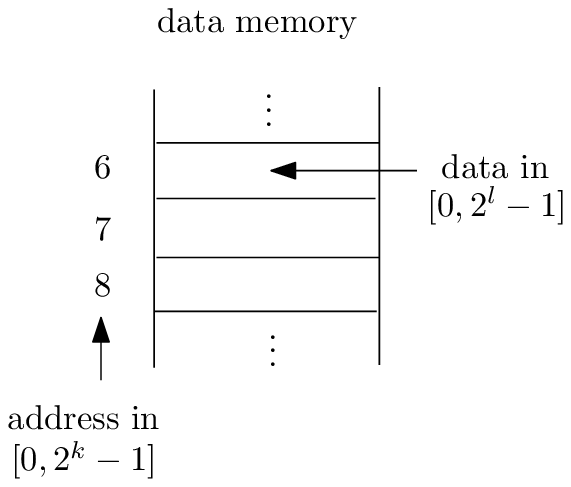}
\caption{Data Memory} \label{figdame}
\end{figure}

The \emph{data memory} is a fixed but arbitrary set $\mdata$ which
has a cardinality of $2^k$ as shown in Figure~\ref{figdame}.
Its elements can contain natural numbers as data in the interval
$[0,2^l-1]$ (written $\bdata$), and can be addressed by natural
numbers in the interval $[0,2^k-1]$ (written $\baddr$). Hence,
we give a fixed but arbitrary bijection $\fmdata
:\baddr\rightarrow\mdata$.

The \emph{operating unit memory} is a fixed but arbitrary set
$\mou$ which has a cardinality of $m$. Its elements can contain
natural numbers in the set $\{0,1\}$ (written $\bit$), i.e.,
bits.

Registers are used to move data between the data memory and the
operating unit memory. \emph{Load address registers} and
\emph{load data registers} are fixed but arbitrary sets $\mla$ and
$\mld$ respectively, which have cardinality of $u$. \emph{Store
address registers} and \emph{store data registers} are fixed but
arbitrary sets $\msa$ and $\msd$ respectively, which have
cardinality of $v$. The contents of $\mla$ and $\msa$ are taken as
addresses which are the members of $\baddr$, while the contents of
$\mld$ and $\msd$ are taken as data which are the members of
$\bdata$. Hence, written $[0,u-1]$ and $[0,v-1]$ as $\setu$ and
$\setv$ respectively, we give fixed but arbitrary bijections
$\fmld:\setu\rightarrow\mld$, $\fmla:\setu\rightarrow\mla$,
$\fmsd:\setv\rightarrow\msd$ and $\fmsa:\setv\rightarrow\msa$.

The memory element $\rr$ stores the reply of processing $\operation_a$, the
operation associated with the basic action $a$.

\begin{figure}[!ht]
\centering
\includegraphics{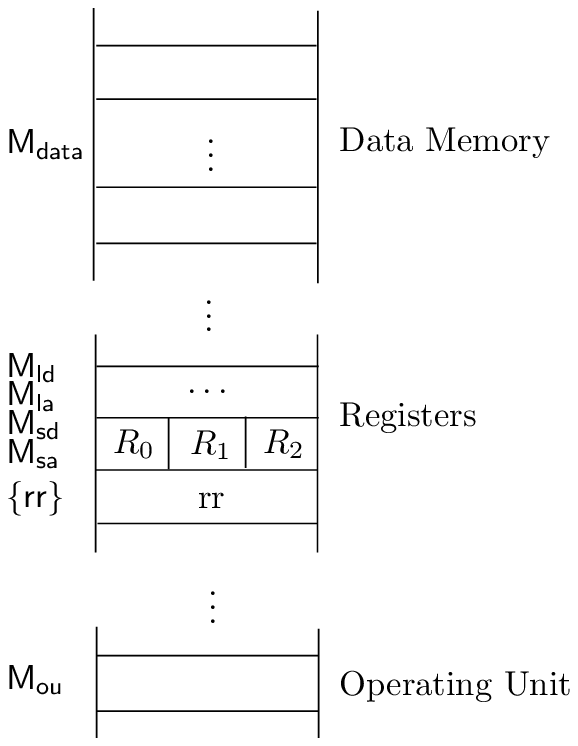}
\caption{The Set Indications} \label{figset}
\end{figure}

We assume that $\mdata$, $\mou$, $\mld$, $\msd$, $\mla$, $\msa$
and $\{\rr\}$ are pairwise disjoint sets.
The meaning of these sets in reality are shown in Figure~\ref{figset}.
Let $n\in\baddr$, $n'\in\setu$ and $n''\in\setv$. Then $\fmdata(n)$ is
denoted by $\mdata[n]$, $\fmld(n')$ by $\mld[n']$, $\fmla(n')$ by
$\mla[n']$, $\fmsd(n'')$ by $\msd[n'']$ and $\fmsa(n'')$ by
$\msa[n'']$.

We give the following definition of a strict load-store Maurer
ISA.
\begin{defn}
\emph{A strict load/store Maurer ISA} with parameters $k$, $l$, $m$,
$u$, $v$ and $\bidm$ is a Maurer machine
$\machine=(\memory,\baseset,\sts,\opns,\bactions,\baif{\_})$ with
\begin{eqnarray*}
\memory&=&\mdata\cup\mou\cup\mld\cup\msd\cup\mla\cup\msa\cup\{\rr\},\\
\baseset&=&[0,j]\cup\blset \textrm{ for } j=\max(2^k-1,2^l-1),\\
\sts&=&\{\state{:} \memory\rightarrow \baseset\mid\\
&&\forall m\in\mdata\cup\mld\cup\msd. \state(m)\in\bdata\land\\
&&\forall m\in\mla\cup\msa.\state(m)\in\baddr\land\\
&&\forall m\in\mou.\state(m)\in\bit\land\state(\rr)\in\blset\},\\
\opns&=&\{\operation_a\mid a\in\bactions\},\\
\bactions&=&\{\loa{:}n\mid n\in\setu\}\cup\{\sto{:}n\mid n\in\setv\}\cup\bidm,\\
\baif{a}&=&(\operation_a,\rr)\textrm{ for all } a\in\bactions,
\end{eqnarray*}
where for all $n\in\setu$, $\operation_{\loa{:}n}$ is the unique
function from $\sts$ to $\sts$ such that for all $\state\in\sts$:
\begin{eqnarray*}
\operation_{\loa{:}n}(\state)\upharpoonright(\memory\setminus \{\mld[n],\rr\})&=&\state\upharpoonright(\memory\setminus \{\mld[n],\rr\}),\\
\operation_{\loa{:}n}(\state)(\mld[n])&=&\state(\mdata[\state(\mla[n])]),\\
\operation_{\loa{:}n}(\state)(\rr)&=&\tr,
\end{eqnarray*}
and, for all $n\in\setv$, $\operation_{\sto{:}n}$ is the unique
function from $\sts$ to $\sts$ such that for all $\state\in\sts$:
\begin{eqnarray*}
\operation_{\sto{:}n}(\state)\upharpoonright(\memory\setminus \{\mdata[\state(\msa[n])],\rr\})&=&\state\upharpoonright(\memory\setminus \{\mdata[\state(\msa[n])],\rr\}),\\
\operation_{\sto{:}n}(\state)(\mdata[\state(\msa[n])])&=&\state(\msd[n]),\\
\operation_{\sto{:}n}(\state)(\rr)&=&\tr,
\end{eqnarray*}
and, for all $a\in\bidm$, $\operation_a$ is a function from $\sts$
to $\sts$ such that:
\begin{eqnarray*}
IR(\operation_a)&\subseteq&\mou\cup\mld,\\
OR(\operation_a)&\subseteq&\mou\cup\msd\cup\mla\cup\msa\cup\{rr\}.
\end{eqnarray*}
\end{defn}

We denote the set of all strict load-store Maurer ISAs with
parameters $k$, $l$, $m$, $u$, $v$ and $\bidm$ by
$\misa(k,l,m,u,v,\bidm)$.
%
%%%%%%%%%%%%%%%%%%%%%%%%%%%%%%%%%%%
\section{Thread Powered Function Classes}\label{secttpf}
In this section we review the thread powered function classes,
which help to answer the following question: under which conditions can we achieve all the possible state
transformations by applying threads to a strict load/store Maurer
ISA with certain address width and word length?

A thread powered function class has the following parameters:
\begin{enumerate}
\item [-] an address width $k$; \item [-] a word length $l$; \item
[-] an operating unit size $m$; \item [-] an instruction set size
$d$; \item [-] a state space bound $e$; \item [-] a working area
flag $f$.
\end{enumerate}
The symbols can be regarded as follows:
\begin{enumerate}
\item [-] $d$: the number of basic instructions excluding load and
store instructions; \item [-] $e$: a bound
on the number of states of the threads that can be applied; \item
[-] $f$: indicates whether a part of the data memory is taken as a
working area. There are two cases. First, if $f=\tr$, we use the
first half of the data memory as the \emph{external memory} and
the second half of the data memory as the \emph{internal data
memory}. Second, if $f=\fa$, we use the whole data memory as the
external memory.
\end{enumerate}

The definition of the thread powered function class is given as
follows.
\begin{defn}
Let $k,m\ge 0$ and $l,d,e>0$, and let $f\in\blset$ such that
$f=\fa$ if $k=0$. We define
\begin{eqnarray*}
\mdata^k &=& \{\fmdata(i)\mid i\in[0,2^k-1]\},\\
\sdata &=& \{\state\mid\state:\mdata^k\rightarrow\bdata\},\\
\tdata &=& \{\tp\mid\tp:\sdata\rightarrow\sdata\}.
\end{eqnarray*}
Then the \emph{thread powered function class} with parameters
$k,l,m,d,e,f$, denoted by $\tpfc(k,l,m,d,e,f)$, which is a
subset of $\tdata$, is defined as follows:
\begin{eqnarray*}
&&\tp \in \tpfc(k,l,m,d,e,f)\\
&&\quad\Leftrightarrow  \exists \bidm \subseteq \bas.\\
&&\quad\quad\exists\machine\in\misa(k,l,m,u,v,\bidm). \\
&&\quad\quad\exists p\in\frterms(\bactions_\machine).\\
&&\quad\qquad (card(\bidm)=d\land card(Res(p))\le e \land \\
&&\qquad\qquad\forall\state\in\sts_\machine. \\
&&\qquad\qquad\quad (( f =\fa \Rightarrow \tp ( \state \upharpoonright \mdata^k)=(p\aply\state)\upharpoonright\mdata^k)\land \\
&&\qquad\qquad\qquad (f=\tr\Rightarrow
\tp(\state\upharpoonright\mdata^k)\upharpoonright\mdata^{k-1}=(p\aply\state)\upharpoonright\mdata^{k-1}))).
\end{eqnarray*}
\end{defn}

Threads are stored in the data memory. When the internal data memory
is used as a part of the operating unit, threads are stored in the
external memory.

We say that $\tpfc(k,l,m,d,e,f)$ is \emph{complete} if $\tpfc(k,l,m,d,e,f)$
is equal to $\tdata$.

The following theorem points out that we can get the completeness if
we use 5 data manipulation instructions and threads with at most
$6+w$ states ($w$ is the number of load and store instructions) and
take the operating unit size slightly greater than the data memory
size.

The 5 data manipulation instructions (recall that load and store
instructions are not counted for the instruction set) are as
follows: an initialization instruction, a pre-load instruction, a
post-load instruction, a pre-store instruction, and a transformation
instruction. First, before a data memory element $m_0$ is moved to
any register, the address of $m_0$ is sent to the load address
register by the pre-load instruction. And then $m_0$ is loaded to
the load data register. Next, the post-load instruction moves the
content of the load data register to the operating unit. Similarly,
before the data is moved from the register to the data memory, the
pre-store instruction sends the intended address in the data memory
to the store address register. And then the content of the operating
unit is moved to the store data register. Next, the content of the
store data register is stored to the data memory. The transformation
instruction applies the relevant state transformation to the content
of the operating unit.

The number of the states of the threads
consists of 5 states associated with the above 5 data manipulation
instructions, the $w$ states associated with load and store
instructions, and the termination state.
\begin{thm}\label{thmcom}
Let $k\ge 0$, $l>0$ and $f\in\{\tr,\fa\}$, and let $dms$ be the data
memory size, i.e., $dms=2^k\cdot l$. Then
$\tpfc(k,l,dms+k+1,5,6+w,f)$ is complete.
\end{thm}

In~\cite{BM:ousl:2007}, a
proof of the case that there are only one load and one store
instructions is given.

The following corollary points out that we can still get the
completeness if we use about half of the data memory size as the
operating unit size.
\begin{cor}\label{corcom}
Let $k,l>0$, and let $ems$ be the external memory size in the case
that $ems$ is half of the data memory size, i.e., $ems=2^{k-1}\cdot
l$. Then $\tpfc(k,l,ems+k,5,6+w,\tr)$ is complete.
\end{cor}

In the cases of Theorem~\ref{thmcom} and Corollary~\ref{corcom}, we need at least 5 data manipulation instructions to accomplish the job. 
%
%%%%%%%%%%%%%%%%%%%%%%%%%%%%%%%%%%%
\section{Incompleteness}\label{sectinc}
In this section we show under which conditions it is impossible to
achieve all transformations on the states of the external memory
taking into account the use of the internal data memory.

\begin{figure}[!ht]
\centering
\includegraphics{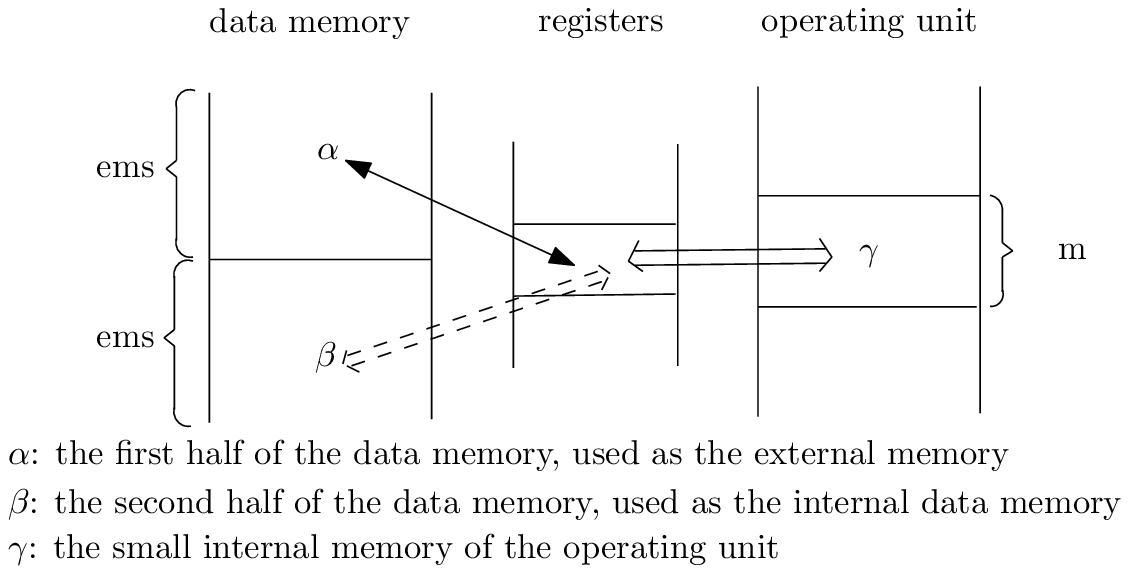}
\caption{Using the Internal Data Memory} \label{fighalf}
\end{figure}

The idea of using the internal data memory can be explained in Figure~\ref{fighalf}.
We move data from $\alpha$ to registers, operate them (e.g., adding
two numbers) in $\gamma$, and then
move the result back to registers. If it is not possible to process
all the operations in $\gamma$ due to the lack of space, we
use $\beta$ and $\gamma$ together to process operations.

In~\cite{BM:ousl:2007}, $\beta$ is not used to process operations in
the case of the lack of space. Lemma 1 in~\cite{BM:ousl:2007} states
that if the operating unit size is at most $ems/2$, the instruction
set size is at most $2^{ems/2}$, and the number of threads that can
be applied is at most $2^{ems}$, it is impossible to achieve all
transformations on the states of the external memory, where $ems$
(external memory size) is half of the data memory size.

We reformulate this lemma with the use of the internal data memory
as follows. It states that it is still impossible to achieve all
transformations on the states of the external memory if the total
size of the operating unit and the used internal data memory is at
most $ems/2$.

\begin{lem}
\label{lemip} Let $k>1$, $l,m,d,e>0$ and $ems=(2^k\cdot l)/2$, and
let $ims$ be the used internal data memory size. Then
$\tpfc(k,l,m,d,e,\tr)$ is not complete if $m+ims\le ems/2$, $d\le
2^{ems/2}$, the number of threads that can be applied to the members
of
\[\bigcup_{\bidm\subseteq\bactions}\misa(k,l,m,u,v,\bidm)\]
is at most $2^{ems}$.
\end{lem}
\begin{proof}
We know that, if the total size of the operating unit and the used
internal data memory is at most $ems/2$, then the number of bits the
operating unit and the used internal data memory have is at most
$ems/2$. As shown in Figure~\ref{figbc},
\begin{figure}[!ht]
\centering
\includegraphics{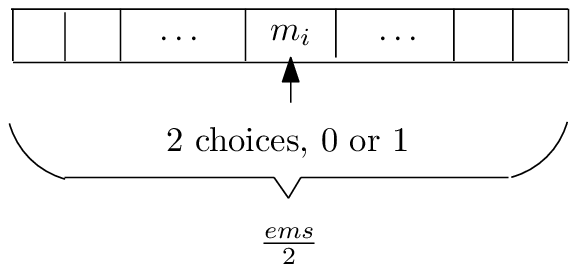}
\caption{Bits in the Memory} \label{figbc}
\end{figure}
since every bit $m_i$ has two choices, $0$ or $1$, for $1\le i\le
ems/2$, the number of states of the operating unit and the used
internal data memory (in other words, the number of sequences that
$ems/2$ digits can make up if every digit has 2 choices) is at most
$2^{(ems/2)}$. Hence there are at most
\[(2^{\frac{ems}{2}})^{(2^{\frac{ems}{2}})}\]
transformations on the states of the operating unit and the used
internal data memory for one data manipulation instruction.

It follows that, if there are at most $2^{ems/2}$ data manipulation
instructions, then there are at most
\[\big((2^{\frac{ems}{2}})^{(2^{\frac{ems}{2}})}\big)^{(2^{\frac{ems}{2}})}\]
transformations on the states of the external memory for one
thread.

So, if at most $2^{ems}$ threads can be applied, then the number of
transformations on the states of the external memory is at most
\[ \big((2^{\frac{ems}{2}})^{(2^{\frac{ems}{2}})}\big)^{(2^{\frac{ems}{2}})}\cdot 2^{ems}.\]

This number is less than the number of all possible transformations
on the states of the external memory, which is
$(2^{ems})^{(2^{ems})}$, i.e.,
\begin{equation}\label{eqnless}
\tag{$\ast$}
\big((2^{\frac{ems}{2}})^{(2^{\frac{ems}{2}})}\big)^{(2^{\frac{ems}{2}})}\cdot
2^{ems} < (2^{ems})^{(2^{ems})}.
\end{equation}
Therefore, we get that $\tpfc(k,l,m,d,e,\tr)$ is not
complete.

We prove (\ref{eqnless}) by the following computation: Let
$x=2^{(ems/2)}$. Then
\begin{align}\label{noneqn}
&&(\ast)\Rightarrow (x^x)^x\cdot x^2 < (x^2)^{(x^2)} \Rightarrow\nonumber \\
&& x^{(x^2)}\cdot x^2 < (x^2)^{(x^2)} \nonumber\Rightarrow \\
&& x^{(x^2)} <  (x^2)^{(x^2-1)} \tag{$\star$}
\end{align}
Applying logarithm to both sides of (\ref{noneqn}), we have
\[x^2\log_2x < 2(x^2-1)\log_2x\Rightarrow (x^2-2)\log_2x>0. \]
If $x>\sqrt{2}$, then we have $x^2>2$, i.e., $x^2-2>0$. Since
$\log_2x
> 1/2$ if $x>\sqrt{2}$, $(x^2-2)\log_2x>0$ holds if $x>\sqrt{2}$,
i.e., $ems>1$.
\end{proof}

Now we can give the following theorem showing that if the total size
of the operating unit and the used internal data memory is at most
$ems/2$, the instruction set size is at most $2^l-w-1$, the maximal number of states
of the threads is at most $2^{k-2}$, then $\tpfc(k,l,m,d,e,\tr)$ is
not complete.
\begin{thm}
\label{thmip} Let $k>2$, $l>1$, $m,d>0$, $e>1$ and $ems=(2^k\cdot l)/2$,
and let $ims$ be the used internal data memory size and $w$ the
number of load and store instructions. Then $\tpfc(k,l,m,d,e,\tr)$
is not complete if $m+ims\le ems/2$, $d\le 2^l-w-1$, $e\le 2^{k-2}$.
\end{thm}
\begin{proof}
We have $d$ data manipulation instructions, plus $w$ load and store
instructions, then there are $d+w$ instructions. Suppose every state of threads can perform either according to the positive reply produced by the associated instruction, or according to the negative reply. Since $e$ is the maximal number of states of the threads that can be applied, no matter which path it performs, the number of states of each path is at most $e$. Hence, we have $d+w$ choices for instructions, $e$ choices for the path caused by the positive reply, and $e$ choices for the path caused by the negative reply. Including the
termination and deadlock, we have $(d+w)\cdot e^2+2$ choices to form a thread. Therefore, the number of threads with $e$ states is
\[\big((d+w)\cdot e^2+2\big)^e.\]
Since $k>2$, $l\ge 2$, $e>1$, we have
\[\big((d+w)\cdot e^2+2\big)^e<\big((d+w)\cdot e^2+e^2\big)^e \le 2^{ems} \textrm{ if } l\ge 2k-4.\]
Hence, the number of threads with $e$ states is less than
$2^{ems}$.

It is easy to see that $2^l<2^{l\cdot 2^{k-2}}=2^{ems/2}$. Then we can get $2^l-w-1<2^{ems/2}$, i.e., $d<2^{ems/2}$. Because $m+ims\le ems/2$, applying Lemma~\ref{lemip},
we can conclude $\tpfc(k,l,m,d,e,\tr)$ is not complete if $m+ims\le ems/2$, $d\le 2^l-w-1$, $e\le 2^{k-2}$.
\end{proof}
%
%%%%%%%%%%%%%%%%%%%%%%%%%%%%%%%%%%%
\section{Conclusion}\label{sectcon}
We have reviewed the concepts of $\bta$ and strict load-store Maurer
ISA. We also have shown under which conditions we can achieve
all the possible transformations on the states of the external
memory of a strict load-store Maurer ISA and under which conditions
we cannot.

From Theorem~\ref{thmcom} and Corollary~\ref{corcom}, we can get
completeness with 5 data manipulation instructions and at most $6+w$
states of the threads if we take the operating unit size slightly
greater than the data memory size, or half of the data memory size.
The completeness is lost by decreasing the number of data
manipulation instructions and the number of states of the threads.
Theorem~\ref{thmip} stated that it is impossible to achieve all
transformations if the total size of the operating unit and the used
internal data memory is at most half of the external memory size,
the instruction set size is at most $2^l-w-1$, and the maximal number
of states of the threads is at most $2^{k-2}$.
%
%%%%%%%%%%%%%%%%%%%%%%%%%%%%%%%%%%%
\bibliographystyle{plain}
\bibliography{biblio}
\end{document}